\documentstyle[epsf,mnras_cite,times]{mn}
\input{epsf}
\newcommand{\ba}{\begin{eqnarray}}
\newcommand{\ea}{\end{eqnarray}}
\newcommand{\be}{\begin{equation}}
\newcommand{\ee}{\end{equation}}
\newcommand{\nn}{\nonumber \\}
\newcommand{\vk}{{\bf{k}}}

\newcommand{\vx}{{\bf{x}}}
\newcommand{\ls}{\mathrel{\raise1.16pt\hbox{$<$}\kern-7.0pt 
\lower3.06pt\hbox{{$\scriptstyle \sim$}}}}         
\newcommand{\gs}{\mathrel{\raise1.16pt\hbox{$>$}\kern-7.0pt 
\lower3.06pt\hbox{{$\scriptstyle \sim$}}}}         
\def\VEV#1{{\langle #1 \rangle}}
\long\def\comment#1{}
\def\hatn{{\bf \hat n}}
\def\fun#1#2{\lower3.6pt\vbox{\baselineskip0pt\lineskip.9pt
  \ialign{$\mathsurround=0pt#1\hfil##\hfil$\crcr#2\crcr\sim\crcr}}}
\def\lap{\mathrel{\mathpalette\fun <}}
\def\gap{\mathrel{\mathpalette\fun >}}
\title{Large-scale structure, the cosmic microwave background, and
primordial non-gaussianity} 
\author[L. Verde, L. Wang, A.F. Heavens,  M. Kamionkowski ]
{Licia Verde$^{1}$, Limin  Wang$^{2}$, Alan F. Heavens$^{1}$ and
Marc Kamionkowski$^{2}$\\
$^{1}$ Institute for Astronomy, University of Edinburgh, Royal Observatory, 
Blackford Hill, Edinburgh EH9 3HJ, United Kingdom\\
$^{2}$ Department of Physics, Columbia University, 538 West 120$^{th}$ Street,
New York, NY 10027, USA\\}
\begin{document}
\maketitle
\begin{abstract}
Since cosmic-microwave-background (CMB) and
large-scale-structure (LSS) data will shortly improve
dramatically with the
Microwave Anisotropy Probe (MAP) and Planck Surveyor, and 
the Anglo-Australian 2-Degree Field (2dF) and Sloan Digital Sky
Survey (SDSS), respectively, it is timely to ask which of the CMB
or LSS will provide a better probe of primordial non-gaussianity.  In
this paper we consider this question, using the bispectrum as a
discriminating statistic.  We consider several non-gaussian
models and find that in each case the CMB will provide a better
probe of primordial non-gaussianity.  Since the bispectrum is the
lowest-order statistic expected to arise in a generic
non-gaussian model, our results suggest 
that if CMB maps appear gaussian, then apparent
deviations from gaussian initial conditions in galaxy surveys can
be attributed with confidence to the effects of biasing. We demonstrate this
precisely for the spatial bispectrum induced by local non-linear biasing.
\end{abstract}
%
%
\section{Introduction}%
It is widely accepted that the large-scale structures we observe 
in the Universe today
originated from gravitational evolution of small primordial
fluctuations in the matter density.  Information about the
physical processes that
generated these primordial fluctuations can be gleaned by testing
whether their statistical distribution is well approximated by a
gaussian random field.  In particular, the simplest versions of
inflation predict gaussian initial fluctuations (e.g.,
\pcite{Guthpi82}; \pcite{Hawking82}; \pcite{Starobinsky82}; \pcite{BST83}), but
there are other models of inflation (\pcite{allengrinsteinwise87};
\pcite{kofpog88}; \pcite{SBB89}) and models where structure is seeded
by topological defects (\pcite{vilke85}; \pcite{vach86};
\pcite{hillscrammfry89}; \pcite{Tur89}; \pcite{albsteb92}) that
generate non-gaussian fluctuations.  
By looking at cosmic-microwave-background (CMB) anisotropies we
can probe cosmic fluctuations at a time when their statistical
distribution should have been close to their original form.  At
present, the limited signal-to-noise or sky coverage of existing
experiments is not sufficient to provide conclusive evidence
either for or against non-gaussianity 
(e.g., \pcite{Hea98}; \pcite{Ferreira98}; \pcite{KamJaf98};
\pcite{PanValFan98}; \pcite{BroTeg99}).  An alternative approach is
to analyze the present-day statistics of density or velocity fields of
large scale structure (LSS).  In principle, this is a more complicated
approach, since gravitational instability and bias can introduce
non-gaussian features in an initially gaussian field, and these may
mask the signal we desire to measure.  Since the CMB and LSS data will
shortly improve dramatically with the Microwave Anisotropy Probe
(MAP) and Planck Surveyor satellites, and
the Anglo-Australian 2-Degree Field (2dF) and Sloan Digital Sky
Survey (SDSS), respectively, it is timely to ask which of the CMB
or LSS is the better place to look to detect a primordial
non-gaussian signal.

In this paper, we use the skewness and bispectrum to
determine which of the CMB or LSS provides a better probe of
non-gaussianity.  To do so, we consider several models with
a primordial non-gaussianity whose amplitude can be dialed from
zero (the gaussian limit).  We then calculate the smallest
non-gaussian amplitude that can be detected with a plausible CMB
map and with  a plausible galaxy survey.  In each case, we find
that the smallest non-gaussian amplitude detectable with a CMB
map is smaller than that from a galaxy survey, even if we
neglect the complicating effects of biasing.

Of course, there is an infinitude of possible deviations from
gaussianity and we cannot address them all.  However,
physical mechanisms that produce non-gaussianity generically
produce a non-vanishing bispectrum.  Since this is the
lowest-order non-gaussian statistic, it is usually the most
easily detectable.  Although this argument is not fully general, 
we show that it holds in several non-gaussian models that have
been considered in the recent literature.  Our results suggest
that if the CMB maps provided
by MAP and Planck are consistent with gaussian initial
conditions, then any signatures of non-gaussianity found in
galaxy surveys (apart from those from non-linear clustering) can
be attributed to biasing.

\section{The skewness}
In order to determine whether a field is gaussian, we need a
discriminating statistic.  We shall principally be concerned with the
bispectrum,  as it is the lowest-order non-gaussian statistic
that generically arises in physical mechanisms that produce
non-gaussianity, and  it is able in principle to distinguish
between various sources of non-gaussianity (e.g. primordial,
non-linear growth, bias).  However, we begin by discussing the
skewness in LSS and the CMB, as this is simpler, and illustrates
some of the effects.

\subsection{Skewness in Large scale structure}
The statistical properties of the fluctuations in the cosmological
mass density field $\delta({\bf x})=[\rho({\bf
x})-\overline{\rho}]/\overline{\rho}$ can be characterized by
the $n$-point moments, $\langle \delta^n\rangle$.
By definition,
$\langle\delta({\bf x})\rangle=0$. If the fluctuation field is
gaussian, then the probability distribution for $\delta$ is
\be
p(\delta)=\frac{1}{\sqrt{2 \pi}\sigma}
\exp\left[-\frac{\delta^2}{2 \sigma^2}\right],
\ee
{}from which the moments ($n=0,1,\ldots$) can be calculated to be
\be
\langle \delta^{2n}\rangle=(2
n-1)!!\langle\delta^2\rangle^n=(2n-1)!!\sigma^{2n},
\label{eq:gaussdelta}
\ee
where $\sigma^2\equiv\langle\delta^2\rangle$.  The odd moments
are of course zero.
To linear order in perturbation theory, $\delta$ grows
by an overall normalization factor, $\delta({\bf x},t)=D(t)\delta({\bf
x},t_0)$, so an initially gaussian distribution will
remain gaussian as long as linear perturbation theory holds.

To higher order in perturbation theory, gravitational
instability will induce departures from gaussianity.  To
describe the evolution of non-linear fluctuations in perturbation
theory we expand the fluctuation field in a series,
\be
\delta=\delta^{(1)}+\delta^{(2)}+\cdot \cdot \cdot,
\ee
where $\delta^{(n)}\!\! \sim\! {\cal O}(\delta^n)$ (e.g. \pcite{Goroff86};
\pcite{frysher94}).
In the weakly non-linear regime, the series can be truncated to second
order.  The lowest-order deviation from gaussianity is described by
$\langle \delta^3\rangle$.  
In the weakly non-linear regime, gaussian initial conditions
give rise to a non-vanishing skewness (e.g. \pcite{Peebles}),
\be
S_3\equiv\frac{\langle\delta^3\rangle}{\langle
\delta^2\rangle^2} = {34 \over 7},
\label{eq:skew}
\ee
in second-order perturbation theory (2OPT).

For generic non-gaussian initial conditions, \scite{frysher94}
found the skewness in second-order perturbation theory (2OPT) to 
be
\be
S_3\!=\!S_{3,0} + \frac{34}{7} -
\frac{26}{21}\frac{\langle\delta^3_0\rangle^2}{\langle\delta^2_0\rangle^3} 
- \frac{8}{7} \langle\delta^3_0\rangle \frac{I[\xi_{(\!3\!)0}\!]}
{\langle\delta^2_0\rangle^3}\! +\! \frac{10}{7}\! \frac{\xi_{(\!4\!)0}}
{\langle\delta^2_0\rangle^2} + \frac{6}{7} \frac{I[\xi_{(\!4\!)0}\!]}
{\langle\delta^2_0\rangle^2},
\label{ngskewness}
\ee
where the subscript $0$ denotes the quantity linearly evolved from the
initial density field [e.g., in an Einstein-de Sitter Universe,
$\delta_0=\delta(z_{i})(1+z_{i})$] to the present epoch.
The quantity $I[\xi_{(n)0}]$ ($n=3,4$) is an integral that
depends on the specific linearly-evolved connected (or irreducible) three- and
four-point functions, $\xi_{(3)0}$ and $\xi_{(4)0}$,
respectively.  (For gaussian initial conditions, $\xi_{(n)0}=0$
for $n\geq3$.)
\scite{frysher94} find that $I[\xi_{(n)0}] \le \xi_{(n)0}(0)$
for $n=3,4$ for several non-gaussian models they explore.
All terms in the RHS of equation (\ref{ngskewness}) are
time-independent apart from $S_{3,0}$ which scales like
$S_{3,0}(z)\propto S_{3,0}(z=0)(1+z)$ in an Einstein-de Sitter
Universe.

It is therefore useful to define the time-independent quantities
$p_3=\xi_{(3),0}/\sigma^3$ (normalized  skewness) and
$p_4=\xi_{(4),0}/\sigma^4$ (normalized kurtosis). 
When written it terms of the relevant quantities at decoupling
(i.e., at $z\simeq1100$),
\be
     S_3 \simeq \frac{34}{7} + \frac{p_3}{1100\,\sigma(z=1100)} + d_1 p_3^2 + 
     d_2 p_4,
\label{secondngskewness}
\ee
where $d_1\simeq d_2 \simeq 2$.

Now suppose we have a survey of $N$ independent volumes in which 
the rms fractional density contrast is $\sigma$.  Then, from
equations (\ref{eq:gaussdelta}) and (\ref{eq:skew}), it follows
that in the mildly non-linear regime, the standard error due to
cosmic variance with which the skewness can be recovered is
\be
 \Delta S_3=\frac{1}{\sqrt{N}}\sqrt{\frac{15}{\sigma^2}+17 S_3^2},
\label{errorske}
\ee
where we have used $\Delta\langle \delta^3 \rangle\!=
\!\sqrt{15\sigma^6+10\langle\delta^3\rangle^2}$ and 
$\Delta\langle \delta^2 \rangle\!\!\sim\!\!\sqrt{2\sigma^4}$.
Of course, since the
mass will be traced by discrete objects (i.e., galaxies), the
shot noise may increase the error estimate in equation
(\ref{errorske}).

So now let us consider the 2dF and/or SDSS.  The present-day
skewness in volumes of side 10 $h^{-1}$ Mpc could be measured
with a standard error at least $\Delta S_3 \sim 20\,N^{-1/2} \sim
10^{-1}$, where $N$ is the number of such cubes in the
survey volume.  Thus, equation (\ref{secondngskewness}) tells us
that a primordial normalized skewness on the $10\, h^{-1}$ Mpc
scale could be identified in a statistically-significant manner
in 2dF and/or SDSS only if $p_3$ exceeded about $10^{-2}$.

\subsection{Skewness in the CMB}
For simplicity, consider an Einstein-de Sitter Universe.  Then a 
region of comoving size $10 h^{-1}$ Mpc (where $h$ is the Hubble 
parameter in units of 100 km~sec$^{-1}$~Mpc$^{-1}$) subtends an
angle of $0.1^\circ$.  Now suppose that a full-sky
cosmic-variance-limited 0.1-degree-resolution CMB map (close to
the Planck Surveyor's specifications) finds that the
distribution of temperature fluctuations $\Delta T/T$ is
consistent with gaussian with a variance $\sigma^2 =
\VEV{(\delta  \rho/\rho)^2} \sim 0.1 {(\delta T/T)^2} \sim (10^{-4})^2$
(where $\delta\rho/\rho$ is the fractional density perturbation
at a redshift $z\simeq1100$, when the CMB decouples).  The
largest primordial normalized  skewness
$p_3\equiv\VEV{(\delta\rho/\rho)^3}/\sigma^3$ that would be
consistent with such a map would be $\sqrt{15/N_{\rm pix}} \sim
10^{-3}$, where $N_{\rm pix}\sim10^{7}$ is the number of
$0.1\times0.1$-degree pixels.  We have neglected instrumental
noise and assumed systematic effects will be under control.
Still, a sensitivity to a value as small as $p_3 \sim 10^{-3}$,
or perhaps an order of magnitude larger, is a realistic
expectation of CMB maps.

If we compare this with the nominal smallest normalized skewness
$p_3\sim10^{-2}$ accessible with LSS, it appears that the CMB has,
perhaps an extra order of magnitude in sensitivity.  On the
other hand, there may be systematic effects in both the CMB and
LSS measurements that may affect both estimates, and our
argument was only qualitative.  Thus, we conclude from this
exercise that the CMB and LSS should provide roughly comparable
sensitivity to a primordial skewness on $10\, h^{-1}$ Mpc
scales, with perhaps a slight edge to the CMB.  Since these
heuristic arguments are inconclusive to orders of magnitude as
to which of the CMB and LSS provides a better probe of
primordial non-gaussianity, we now proceed to consider realistic
models more carefully.

\section{CMB and LSS Bispectra}
%

The skewness has the advantage of being far easier to calculate
than the full three-point correlation function, but it does not
contain as much information.  We therefore prefer to investigate
the bispectrum (the three-point function in Fourier space) for
reasons which have been rehearsed before
(e.g., \pcite{MVH97}). In addition, various theoretical models for
structure formation yield directly Fourier-space quantities, so
the bispectrum allows a more straightforward relation between
measurable quantities and theoretical predictions.

\subsection{LSS Bispectrum}

We define the Fourier transform of the fractional overdensity perturbation
by $\delta_{\vk}=\int
\frac{d^3\vx}{(2\pi)^3}\,\delta(\vx)\exp(-i\vk \cdot \vx)$.
The spatial bispectrum $B(\vk_1,\vk_2,\vk_3)$ is defined by
\be
\langle
\delta_{\vk_1}\delta_{\vk_2}\delta_{\vk_3} \rangle =
B(\vk_1,\vk_2,\vk_3) \delta^D(\vk_1+\vk_2+\vk_3) 
\ee
where the angle brackets denotes an ensemble average or, by the ergodic
theorem, the average over a large volume (fair sample), and $\delta^D$ 
is the Dirac delta function.

To second-order in perturbation theory, the bispectrum may be written
as a sum of a primordial part and a part induced by gravitational
instability:
\ba
 &B(\vk_1,\vk_2,\vk_3)  \simeq B_0 (\vk_1,\vk_2,\vk_3) &   \nn
 & +[2 J(\vk_1,\vk_2) P_0(k_1)P_0(k_2)+ cyc.] &\nn 
& +\int d^3\vk_a
[J(\vk_a,\vk_3\!-\!\vk_a) T^{c}(\vk_a,\vk_3-\vk_a,\vk_1,\vk_2)+ cyc.],& 
\ea
where $J(\vk_a,\vk_b)$ is a function almost independent of the
non-relativistic-matter density $\Omega_0$ and cosmological
constant $\Lambda$ (\pcite{BJCP92}; \pcite{BCHJ95};
\pcite{Bernardeau94b}; \pcite{SCFFHM98}; \pcite{KB99}).  Its detailed form
need not concern us here.  Here, $B_0$ is the primordial
bispectrum (which we wish to probe) linearly 
evolved to redshift $z$, $P_0$ is the power spectrum and $T^c$ denotes the connected trispectrum, which is
the Fourier transform of the connected 4-point correlation function.
The last two terms arise from non-linear gravitational instability;
this depends (see e.g. \pcite{CM94a}) on the 4-point function, which
has a disconnected part (present in gaussian fields), and a connected
part, $T^c$ (which is zero for gaussian fields).  In principle,
this last term ($\equiv B_T$)
may be important, as it grows as
fast as the usual disconnected part.  We will show later that this term
is very small for a range of proposed models.

As already seen for the skewness, in an Einstein-de Sitter Universe,
$B_0(z) \propto (1+z)^{-3}$ and $[P_0(z)]^2
\propto (1+z)^{-4}$, so the primordial bispectrum redshifts away in
comparison with that arising from non-linear gravitational clustering
\footnote{In a non-Einstein-de-Sitter model, the suppression
factor $(1+z)$ is replaced by $g(z)(1+z)$, but for
reasonable values of $\Omega_0$ and $\Lambda$, this factor is
$0.3 \lap g(z) \lap 1$.  Thus, we will assume an
Einstein-de Sitter model from now, noting that the general
arguments are essentially unchanged in more general models.}.

If we suppose that the galaxies (subscript $g$) are locally
biased with respect to the mass, then the density of galaxies
$\delta_g$ can be Taylor expanded in the mass density to second order,
$\delta_g = b_0 + b_1\delta+\frac{1}{2}b_2\delta^2$,
for some biasing coefficients $b_i$. The quantity $b_0$ affects
only $\vk={\bf 0}$, so it can be ignored.  The bispectrum for the 
galaxies is then
\ba
& B_g(\vk_1,\vk_2,\vk_3) \simeq b_1^3B_0(\vk_1,\vk_2,\vk_3)+ &
\nn & \left\{P_0(k_1)P_0(k_2)\left[b_1^3 2 J(\vk_1,\vk_2) + b_2b_1^2
\right] 
\right\}+ \mbox{cyc.} +b_1^3 B_T.&
\ea
So, assuming that the initial (or the linear) power spectrum is precisely
known, we can do a likelihood analysis only if we also have a model for the
initial (or linear) bispectrum and trispectrum as a function of
the three wave numbers.
As we found previously, the primordial bispectrum redshifts away, and
we can realistically only detect non-gaussianity if $B_T$ is significant.
In section (4.1) we will  quantify further this argument by
showing that for all the models considered $B_T\ll B_0$.
Let us parameterize the observed bispectrum as
\be
B_g(\vk_1,\vk_2,\vk_2)=P_0(k_1)P_0(k_2)[c_1
2J(\vk_1,\vk_2)+c_2]+cyc.
\label{eq:Bg}
\ee
In order to asses the precision with which the bias could be measured,
 \scite{MVH97} evaluated the precision with which the parameters
$c_1$ and $c_2$ could be recovered from a likelihood analysis of 
the 2dF and/or SDSS.  From their Figure 7, we conclude that if
the positions of all of the galaxies in the survey volume are
known, then (a)  $c_1$ can be determined with an error of
$\sim6\times 10^{-3}$ if $c_2$ is fixed; (b) $c_2$ can be
determined with an error of $\sim1\times 10^{-2}$ if $c_1$ is
fixed; and (c) the joint determination of the two parameters
allows $c_1$ and $c_2$ to be recovered with errors of  $1\times
10^{-2}$ and  $4\times 10^{-2}$, respectively.  The analysis of
\scite{MVH97} further shows that the bispectrum signal comes
primarily from $k\simeq0.1-1 \, h$~Mpc$^{-1}$. (All of these
estimates would  be reduced only by about a factor 20 if we
could map the mass throughout the entire Hubble volume rather
than just in the survey volume.) We will use these results to asses the
 smallest detectable primordial non-gaussianity.

\subsection{CMB Bispectrum}

A CMB map of the temperature $T(\hatn)$ as a function of
position $\hatn$ on the sky can be decomposed into spherical
harmonics,
\be
     \frac{\Delta T(\hatn)}{T} = \sum_{lm} a_{lm} Y_{lm}(\hatn),
\ee
where the multipole coefficients are given by the inverse
transformation,
\be
     a_{lm} = \int\, d\hatn \, Y_{lm}^*(\hatn) \frac{\Delta
     T(\hatn)}{T}.
\ee
The CMB bispectrum $B_{l_1 l_2 l_3}$ is defined by
\be
\langle
a_{l_1}^{m_1}a_{l_2}^{m_2}a_{l_3}^{m_3}\rangle=
B_{l_1 l_2 l_3}\left(\begin{array}{ccc}
l_1 & l_2 & l_3 \\
m_1 & m_2 & m_3 \\ 
\end{array}\right).
\ee
where the factorization ensures statistical isotropy, and the last term is the
Wigner 3 J symbol.

If we parametrize the bispectrum by $X$ times some fixed function---as in
equation (14)---of
$l_1,l_2,l_3$,  then the error on
$X$ is given by (see, e.g., \pcite{MVH97}) 
\ba
\sigma_{X}^{-2}&=\left<-\frac{\partial^2 ln {\cal L}}{\partial
X^2}\right> \simeq  \sum_{l_1 \le l_2 \le
     l_3} \frac{(B_{l_1l_2l_3})^2}{n C_{l_1}C_{l_2}C_{l_3}}  \nn
       & \times \sum_{m_1,m_2,m_3}\frac{\left(\begin{array}{ccc}
     l_1 & l_2 & l_3 \\
     m_1& m_2 & m_3 \\
\end{array}\right)^2}{N(m_i,l_i)},
\label{apriorierror}
\ea
where ${\cal L}$ denotes the likelihood function and $C_l\equiv \langle \mid
a^m_{l}\mid\rangle$, is the power spectrum of the sky fluctuation.  Here we assumed
that the departures from gaussianity are small and therefore the
covariance matrix can be approximated by the covariance of a gaussian
field that has the same power spectrum (\pcite{Jungman};
\pcite{Hea98}), and we ignore the mixing which arises from partial sky
coverage.  

If one considers only the real part of $\langle
a_{l_1}^{m_1}a_{l_2}^{m_2}a_{l_3}^{m_3}\rangle$ (see \pcite{MVH97})
then $n=1/2$.  The quantity $N(m_i,l_i)$ is related to the
number of non-zero terms like $C_{l_1}C_{l_2}C_{l_3}$ in the
covariance and ranges from $1$ to 30.  Equation
(\ref{apriorierror}) is valid as long as the noise does not
dominate the signal.  If $N=1$, then the sum over the Wigner 3J
symbols is unity; the fact that $N$ is not equal to one reduces
the sum by a only few percent, so
\be
\sigma_{X}^{-2} \sim  2 \sum_{l_1 \le l_2 \le
l_3}\frac{(B_{l_1l_2l_3}|_{X=1})^2}{C_{l_1}C_{l_2}C_{l_3}}.
\label{eq:sigmaX}
\ee
This expression ignores the effects of pixel noise.  However,
the results we show below use $l\lap 100$ only, and for these 
modes, the pixel-noise variance in MAP and Planck will be
negligible compared to the cosmic variance.

\section{Some non-gaussian models}

We now proceed to consider several classes of
physically-motivated models with primordial non-gaussianity in
order to investigate more precisely the relative sensitivities
of the CMB and LSS.  Since current measurements of CMB and LSS
power spectra are roughly compatible with cold-dark-matter (CDM) models for
structure formation, we first
consider CDM-like models and introduce some non-gaussianity in
several different ways.  Specifically, we consider models in
which the gravitational potential contains a part that is the
square of a gaussian random field and models in which the
density  contains a part that is the square of a gaussian random
field.  We note that such non-gaussianity may arise in slow-roll
and/or nonstandard (e.g., two-field) inflation models
(\pcite{Luo94}; \pcite{FRS93}; \pcite{GLMM94};
\pcite{Fanbardeen92}).  Moreover both of these models may be considered as
Taylor expansions of more general fields, and are thus a fairly generic form
of non-gaussianity.  
We also consider $O(N)$-$\sigma$ models, as these will approximate the
non-gaussianity expected in topological-defect models.

\subsection{Quadratic model for the potential}

We start by considering a model in which the gravitational
potential $\Phi$ (in conformal Newtonian gauge) is a linear
combination of a gaussian random field $\phi$ and a term
proportional to the square of the same random field,
\be
\Phi = \phi + \alpha(\phi^2 - \VEV{\phi^2}),
\ee
where $\alpha$ parametrizes the non-gaussianity; in the limit
$\alpha\rightarrow0$, the model becomes gaussian.\footnote{We
note that such non-gaussianity can arise in standard slow-roll
inflation, and the parameter $\alpha$ can be related to
inflaton-potential parameters (e.g.,
\pcite{FRS93}; \pcite{GLMM94}; \pcite{Wang99}).}
These models contain a primordial bispectrum for the
gravitational potential,
\be 
B_{\Phi}(\vk_1,\vk_2,\vk_3)\simeq
2\alpha[P_{\Phi}(k_1)P_{\Phi}(k_2)+cyc.]. 
\label{eq:BPhi}
\ee
The leading terms in the connected trispectrum for
the $\Phi$ field will be
\ba
& T^{c}_{\Phi}(\vk_1,\cdot\cdot\cdot,\vk_4)\simeq & \nn
&\!\!\!\!4\alpha^2\! P_{\Phi}\!(\vk_1)P_{\Phi}\!(\vk_2)
\!\left[P_{\Phi}(\mid\!\vk_1\!+\!\vk_3\!\mid)+\!
P_{\Phi}\!(\mid\!\vk_1\!+\!\vk_4\!\mid)\right]+{\rm cyc.}&
\ea
We use a scale-invariant primordial spectrum, $P_\Phi=A_H
k^{-3}$, where the amplitude is fixed by COBE to be, in our
Fourier transform conventions, $A_H\simeq 10^{-10}$.

\begin{figure}
\begin{center}
\setlength{\unitlength}{1mm}
\begin{picture}(90,55)
\includegraphics{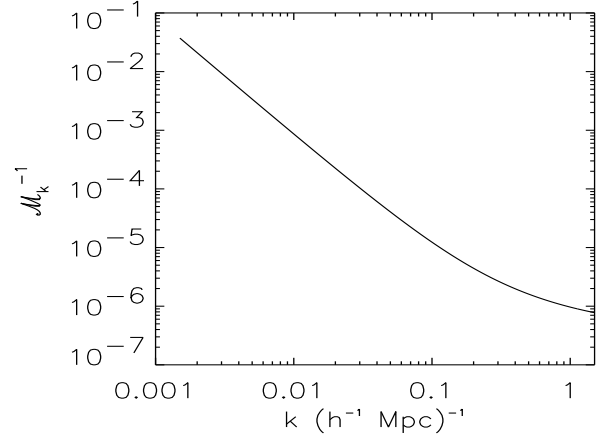}
\end{picture}
\end{center}
\caption{The inverse of the function ${\cal M}_k$ evaluated at
redshift $z=0$.  In the regime where 2OPT holds, this coefficient
is between $10^{-5}$ and $10^{-6}$.}
\label{factor}
\end{figure}

The relative contribution to the bispectrum of $B_0$ and $B_T$,
$R\equiv B_0/B_T$, can be obtained  from equation (9)
and (19). By evaluating the integral we find:
\be
R\sim \frac{1}{\alpha A_H},
\ee
Thus, $B_T \ll B_0$ for $\alpha \ll 10^{10}$, so we can safely neglect the
contribution from the trispectrum.
The $B_T$ contribution to the bispectrum will therefore be
negligible compared with $B_0$ if $\alpha \ll (A_H)^{-1}$.

The Fourier coefficients $\Phi(\vk)$ for the gravitational
potential remain constant to linear order in perturbation theory 
in an Einstein-de-Sitter Universe.  The Fourier coefficients
$\delta(\vk,z)$ of the density field evolve with time and are
related to those of the gravitational potential by the Poisson
equation, which can be written
\be
\delta(k,z) = {\cal M}_k(z)\Phi(k), \mbox{  where   } {\cal M}_k(z)=\frac{2
k^2 T(k)(1+z)}{3 H_0^2}.
\ee
The inverse of the function ${\cal M}_k(z=0)$ is plotted as a
function of wavenumber $k$ in Fig. \ref{factor}.
Thus, the linearly-evolved power spectrum for the mass, $P_0(k,z)$ at
redshift $z$ is related to that for the gravitational potential
(which remains constant in and Einstein-de-Sitter Universe) by
\be
P_0(k,z)=[{\cal M}_k(z)]^2 P_{\Phi}(k),
\ee
where the transfer function is (e.g., \pcite{BBKS})
\be
T(k)\simeq\frac{1}{(1+Bk+Ck^{3/2}+Dk^2)},
\ee
and $B=1.7 \mbox{Mpc}(\Omega h^2)^{-1}$,
$C=9\mbox{Mpc}^{3/2}(\Omega h^2)^{-3/2}$ and $D=1
\mbox{Mpc}^2(\Omega h^2)^{-2}$.  The bispectrum for the mass for
this model is thus
\ba
 & B(\vk_1,\vk_2,\vk_3) \simeq & \nn
&\left\{ P(k_1)P(k_2)\left[\left( 2 \alpha \frac{ {\cal M}_{k_3}}{{\cal
 M}_{k_1} {\cal M}_{k_2} } \right) +2 J(\vk_1,\vk_2)\right]\right\}+cyc. & 
\label{bfull}
\ea

Comparing equation (\ref{bfull}) with equation (\ref{eq:Bg}), we
see that this particular form of primordial non-gaussianity
leads to a present-day bispectrum that looks like a
scale-dependent non-linear bias.  In particular, for equilateral
configurations, equation (\ref{bfull}) becomes identical to
equation (\ref{eq:Bg}) if we identify $c_2=2\alpha/{\cal M}_k$.
We can therefore use the results of \scite{MVH97} to determine
the smallest $\alpha$ that would be detectable by 2dF/SDSS,
under the assumption that there were no non-linear biasing.  In
the range where most of the signal comes from, ${\cal M}_k^{-1}
\simeq 10^{-5} - 10^{-6}$.  Thus, we conclude that the smallest
$\alpha$ that would give rise to an observable signal in the 
2dF/SDSS bispectrum is $\alpha\sim10^3-10^4$.

The primordial gravitational-potential bispectrum in equation
(\ref{eq:BPhi}) will lead to a non-zero bispectrum in the CMB via 
the Sachs-Wolfe effect, and this bispectrum can be calculated to 
be (\pcite{Luo94}; \pcite{Wang99}),\footnote{Actually, our
calculation ignores the physics that gives rise to the acoustic
peaks at $l\gap 100$.  However, we restrict our analysis to
$l\lap 100$.  A more accurate treatment would change our
results by no more than an order of magnitude, and 
this level of precision is sufficient for our
purposes.}
\ba
B_{l_1 l_2 l_3}=\sqrt{\frac{(2 l_1+1)(2l_2+1)(2l_3+1)} {4
\pi}}\left(\begin{array}{ccc}
l_1 & l_2 & l_3 \\
0& 0 & 0 \\
\end{array}\right) \nn
\times \frac{2\alpha}{A_{\rm SW}} \left[ C_{l_1}C_{l_2}+C_{l_1}
C_{l_3}+C_{l_2}C_{l_3}\right],
\label{luobisp}
\ea
where $A_{\rm SW}\simeq1/3$ is the Sach-Wolfe coefficient.
Plugging this bispectrum (and a scale-invariant set of $C_l$)
into equation (\ref{eq:sigmaX}), we learn that the smallest
$\alpha$ that could be detectable with a CMB map (using only
$l\leq100$) is $\sim20$.\footnote{This result is in rough agreement
with the conclusion that the $\alpha\sim10-100$
(\pcite{Luo94}; \pcite{GLMM94}; \pcite{MGLM95}) non-gaussian
signal from non-linear evolution prior to $z\simeq 1100$ is at
best marginally detectable (\pcite{Spergelgoldberg99}).}
(The dashed curve in Fig. \ref{figapriorierror} shows the
smallest $\alpha$ detectable with the CMB as a function of the largest
multipole moment $l$ used in the analysis.)  Thus, we
conclude that the CMB will be at least two orders of magnitude
more sensitive to a non-zero value of $\alpha$ than LSS.

\subsection{Quadratic model for the density}
We now consider an alternative model in which the density field
(rather than the gravitational potential) contains a term that
is the square of a gaussian random field, 
\be
\delta=\phi+\alpha\left(\phi^{2}-\langle\phi^{2}\rangle\right),
\label{eq:deltaphi}
\ee
where now $\phi$ is some other gaussian random field.  Such a
model has been considered in some two-field inflation models
(e.g. \pcite{LS93}).  

Since the density perturbation evolves in
linear theory with time (unlike the gravitational potential), we 
must specify the epoch at which the density perturbation
$\delta$ is related to $\phi$ through equation
(\ref{eq:deltaphi}).  We choose this epoch to be the current
epoch, $z=0$.  Doing so, the spatial mass bispectrum today
due to primordial non-gaussianity is
\be
B_0(\vk_1,\vk_2,\vk_3) \simeq 2 \alpha P(k_1) P(k_2)+cyc.,
\label{eq:deltasquared}
\ee
and \scite{LS93} have shown that the contribution to the
present-day bispectrum from the primordial trispectrum is
negligible compared with this ($B_T \ll B_0$), as long as 2OPT
holds.  By comparing with equation (\ref{eq:Bg}), we see
that this form of primordial non-gaussianity gives rises to a
present-day skewness that mimics precisely that due to
a scale-independent non-linear bias.  Again, from the results of
\scite{MVH97}, the smallest detectable $\alpha$ in this model is 
$\sim0.01$.

\begin{figure}
\begin{center}
\setlength{\unitlength}{1mm}
\begin{picture}(88,60)
\includegraphics{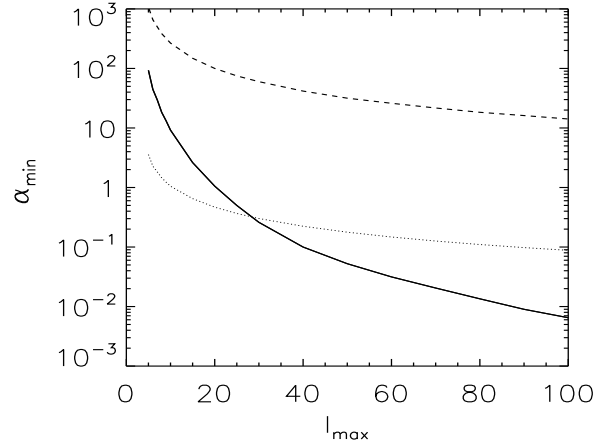}
\end{picture}
\end{center}
\caption{The smallest non-gaussian amplitude $\alpha$ that can
be detected with the CMB as a function of the largest multipole
moment $l$ used in the analysis.
The solid line is for a model in which the present-day fractional overdensity
field contains a term that is the square of a gaussian random
field; the dashed line is for a model in which the gravitational potential
contains a term that is the square of gaussian random field; and
the dotted line
refers to the $O(N)$ $\sigma$ models. }
\label{figapriorierror}
\end{figure}

Now let us consider the CMB bispectrum of this model.  The
spatial bispectrum for the gravitational potential here is
\be
B_{\Phi}(\vk_1,\vk_2,\vk_3)=\frac{2\alpha {\cal M}_{k_1}{\cal M}_{k_2}}{{\cal
M}_{k_3}}\left[P_{\Phi}(k_1)P_{\Phi}(k_2) \right]+cyc.
\ee
Although we have not carried out an exact calculation of the
bispectrum for this model, it is easily seen (at least in the
$l\gg1$ limit) to be
\ba
B_{l_1 l_2 l_3}\simeq\sqrt{\frac{(2 l_1+1)(2l_2+1)(2l_3+1)}{4 \pi}}
\left(\begin{array}{ccc}
l_1 & l_2 & l_3 \\
0& 0 & 0 \\
\end{array}\right) \nn
\times \frac{2\alpha}{A_{SW}} \left[\frac{2}{3} C_{l_1}C_{l_2}
\frac{l_1^2l_2^2}{l_3^2}+cyc.\right].
\label{ansazt}
\ea
Applying equation (\ref{eq:sigmaX}), we find that the smallest
detectable $\alpha$ in this model is $\sim 0.01$ (using only
$l\leq100$; results for other $l_{\rm max}$ are shown in
Fig. \ref{figapriorierror}),
which is comparable to the LSS error.  However,
since this model has more power in the CMB bispectrum at larger
$l$, the smallest detectable $\alpha$ decreases by roughly an
order of magnitude even if we go out only to $l_{\rm max}=200$.
Also in this case we find that CMB will provide a more precise 
probe of a primordial bispectrum.
Moreover, as a corollary,
this particular result demonstrates that if the maps provided by
MAP and Planck are consistent with gaussian, then any
measurement of a non-zero $c_2$ from the LSS bispectrum can be
interpreted unambiguously as evidence for non-linear biasing,
rather than some primordial non-gaussianity.

\subsection{$O(N)$ $\sigma$ models}
The $O(N)$ sigma model provides an approximation to
the nongaussianity expected in topological-defect or
scalar-field-alignment models (\pcite{Turokspergel91};
\scite{Jaffe94}).  The $N=1$ model has domain
walls, $N=2$ global strings, $N=3$ global monopoles, $N=4$
global textures, and higher $N$ correspond to
non-topological-defect models.
For large $N$ this model approaches the gaussian model, so we take
$\alpha=N^{-1/2}$ and as $\alpha \longrightarrow 0$ the models become
asymptotically gaussian.  Since calculation of power spectra and
higher-order statistics for the CMB and LSS are quite involved
for these models, our analysis will be only approximate.  As we
will see below, these order-of-magnitude estimates will be
sufficiently precise for our purposes.

For equilateral-triangle configurations and for values of $k$
that can be probed with 2dF/SDSS, the power spectrum,
bispectrum and trispectrum for LSS in the linear regime are
(\pcite{Jaffe94}) 
\be
P\simeq12.5\,{\cal K}^2 k T^2(k), \quad
B_0\simeq 1.6\, {\cal K}^3 \alpha T^3(k),
\label{sigmaP}
\ee
\be
T^c\simeq {\cal K}^4\frac{1}{k \alpha^2}T^4(k),
\label{sigmatrisp}
\ee
where ${\cal K}\simeq30-100$~Mpc$^2$~$h^{-2}$.
{}From equations (\ref{sigmaP}) and (\ref{sigmatrisp}), $B_T$ is
found to be always $\lap 0.4 B_0$. Since $B_0$ will be the
dominant contribution to the bispectrum we conclude that the non-gaussianity
of this model will act like a scale-dependent non-linear-bias
contribution with
\be
c_2 \simeq \frac{B_0}{P^2}\simeq \alpha[{\cal K} k^2 10^2 T(k)]^{-1}.
\ee
For the scales probed by SDSS/2dF, $ c_2\lap \alpha/300$, and 
therefore the minimum $\alpha$ detectable from LSS will be $\sim30$.

Precise calculation of the CMB bispectrum for these models is
well beyond the scope of this paper.  To obtain an
order-of-magnitude estimate, we assume that the spatial
polyspectra for this model, equations
(\ref{sigmaP})--(\ref{sigmatrisp}), give rise to potential
polyspectra through the Poisson equation and then that the CMB
anisotropy is proportional to the potential perturbation at the
surface of last scatter.  Doing so, we find\footnote{Details
will be presented elsewhere.  Our
estimate differs from that of \scite{Luo94} for reasons that
escape us.  We have checked, however, that our conclusions are
unaltered if we use his results.}
\ba
B_{l_1l_2l_3}\sim\sqrt{\frac{(2 l_1+1)(2l_2+1)(2l_3+1)}{4 \pi}}
\left(\begin{array}{ccc}
l_1 & l_2 & l_3 \\
0& 0 & 0 \\
\end{array}\right) \nn
\times 5000 \, \alpha\left[C_{l1}C_{l_2}C_{l_3}\right]^{2/3}
\ea
Using equation (\ref{eq:sigmaX}), we find that the smallest
detectable $\alpha$ would be $\sim 0.1$ (using multipole moments 
up to $l_{\rm max}=100$; the dependence on $l_{\rm max}$ is
indicated in Fig. \ref{apriorierror}). Thus, we again conclude
that the CMB will provide a more precise probe of a primordial
non-gaussianity.

\section{Conclusions}

We addressed the question of which of the CMB or LSS is better poised
to detect primordial non-gaussianity of several varieties.  We
used the bispectrum as a discriminating statistic since it is
the lowest-order quantity that has zero expectation value for a
gaussian field.  We considered three forms of non-gaussianity: one 
in which the gravitational potential contained a term that was
the square of a gaussian field;  another in which the density 
field was the square of a gaussian field; and a third that resembles that
expected from topological defects.  We showed that in
all cases, the CMB is likely to provide a better probe of such
non-gaussianity.  One of these models produced a mass bispectrum
that mimicked a scale-dependent non-linear bias, and the others
mimicked a scale-independent non-linear bias. 

Of course, our results are not fully general.  In principle, it
is possible to think of some other type of non-gaussianity for
which our conclusions would not hold.  However, plausible
physical mechanisms that produce nearly scale-invariant power
spectra should generically produce non-gaussian signals that have
scale dependences roughly like those that we investigated.  
Thus we may
conclude that if CMB maps turn out to be consistent with
gaussian initial conditions, any non-gaussianity seen in the LSS bispectrum
can be unambiguously attributed to the effects of biasing.

\section*{Acknowledgments}
LV acknowledges the support of TMR grant. LV and AFH thank
the Physics Department of Columbia University for hospitality.
This work was supported at Columbia by a DoE Outstanding
Junior Investigator Award, DE-FG02-92ER40699, NASA grant NAG5-3091,
and the Alfred P. Sloan Foundation.

\end{document}